\documentstyle[aps,preprint,floats,psfig]{revtex}

\begin{document}
\draft
\preprint{\vbox{ \it Submitted to Phys. Rev. C 
\hfill\rm CU-NPL-1158}}

\title{Higher Nucleon Resonances in Exclusive 
$(\gamma, \pi N)$ Reactions on Nuclei}
\author{Frank X. Lee~$^1$~\footnote{Address after 24 August 1998:
Department of Physics,
The George Washington University, \\Washington, DC 20052}, 
Cornelius Bennhold~$^2$, 
Sabit S. Kamalov~$^3$~\footnote{On leave from 
Laboratory of Theoretical Physics, JINR Dubna, SU-101000 Moscow, Russia.},
Louis E. Wright~$^4$}
\address{$^1$~Nuclear Physics Laboratory, Department of Physics,
              University of Colorado, \\ Boulder, CO 80309 \\
$^2$~Center for Nuclear Studies, Department of Physics,
     The George Washington University, \\ Washington, DC 20052 \\
$^3$~Institut f\"{u}r Kernphysik, Johannes Gutenberg-Universit\"{a}t, 
D-55099 Mainz, Germany\\ 
$^4$~Institute of Nuclear and Particle Physics,
     Department of Physics and Astronomy, \\ Ohio University, Athens, OH 45701}
\maketitle

\begin{abstract}
We report calculations for exclusive pion photoproduction 
on nuclei beyond the first resonance region 
in the distorted wave impulse approximation.
The elementary operator contains contributions from the resonances 
$P_{33}(1232)$, $P_{11}(1440)$, $D_{13}(1520)$, $D_{33}(1740)$, 
$S_{11}(1535)$, $S_{11}(1650)$, and $F_{15}(1680)$,
in addition to the usual Born plus vector meson contributions. 
It gives a good description of single pion photoproduction data 
up to about 1.1 GeV.
Final state interactions are incorporated via optical potentials and 
are found to be substantial in the coincidence cross sections, 
but insensitive to the photon asymmetry.
Clear signatures of possible medium modifications of the $D_{13}$ and 
$F_{15}$ resonances are predicted.
\end{abstract}
\vspace{1cm}

\section{Introduction}
\label{intro}

Meson photoproduction has been a valuable 
source of information on the structure of hadrons and nuclei.
With the advent of the new generation of high-intensity continuous-wave 
electron accelerators, as represented by the Jefferson Lab, 
the field reaches a new level of promise. 
The new facilities allow exclusive measurements with unprecedented
precision under a variety of kinematical conditions,
thus making it possible to study higher excited states of hadrons in
greater detail.

The elementary single pion photoproduction (with four possible channels),
\begin{eqnarray}
\gamma + N \rightarrow \pi + N, \label{reaction1}
\end{eqnarray}
has been the subject of extensive study both theoretically and
experimentally over the last two decades.
In the energy range from threshold through the delta resonance region,
a progressive series of models 
based on the effective Lagrangian approach at tree level have been 
developed~\cite{Blomqvist77,Davidson91}
which have generally provided an adequate description of 
available cross section data.
More complete descriptions include final state rescattering by 
iterating the full scattering
equation~\cite{Nozawa90,Sato96,Surya96,Julich96,Feuster98},
thus incorporating unitarity dynamically.
Attempts to extend the tree-level approach beyond the delta resonance region 
have been made in recent years in Ref.~\cite{Hanstein93} and 
Ref.~\cite{Garcilazo93}
for energies up to around $E_\gamma$=1 GeV. 
At even higher energies ($E_\gamma \geq$ 4 GeV), a model for pion and
kaon photo- and electroproduction by 
Guidal {\it et. al.}~\cite{Guidal97} has been developed 
based on Regge theory.
A fairly complete compilation of the data below 2 GeV 
has been maintained by the VPI group~\cite{SAID97} and 
by the KEK group~\cite{Ukai85}.

Using a newly developed operator,
we carry out nuclear calculations in this work 
for the exclusive reactions, 
$A(\gamma, \pi N)A-1$, beyond the first resonance region
in a distorted wave impulse approximation (DWIA) framework. 
Pion photoproduction on a nucleus provides a different environment
in which the reaction can take place. 
Issues arise such as the nuclear reaction 
mechanism, the interaction of the resonances with the medium, and 
the final state interactions.
Our focus is on the puzzle of the `damped resonances'
in the second and third resonance regions as seen in inclusive 
photoabsorption cross section data on various nuclei~\cite{Bianchi93}.
The data show an unexpected damping behavior of the higher 
resonances when compared with the same process on the proton and the deuteron. 
Clearly, in order to isolate the mechanism for this mysterious
phenomenon the individual exclusive channels need to be investigated. 

Quasifree pion photoproduction allows for the study of the production
process in the nuclear medium as well as final state interaction
effects without being obscured by the details of the nuclear transition.
This is due mainly to the quasifree nature
of the reaction which permits the kinematic flexibility to have small
momentum transfers. Conceptually, the initial nucleus is a target holder
which presents a bound nucleon to the incoming photon beam.
The basic reaction $N(\gamma, \pi N)$ takes place in the nuclear medium
producing a continuum pion and nucleon which interact with the residual
nucleus as they exit the target. The key ingredients in such a
description are:
a) the single-particle wave function of the initial nucleon and
spectroscopic factor,  usually taken from electron scattering,
b) the elementary pion photoproduction amplitudes, obtained from
measuring the free processes,
c) the nucleon-nucleus final state interaction, taken from elastic 
scattering,  and finally,
d) the pion-nucleus final state interaction, also taken from elastic 
scattering.

The ability of the reaction in studying these issues 
has been demonstrated in our previous works
on pion photoproduction~\cite{Lee93} and electroproduction~\cite{Lee97}
from nuclei in the delta resonance region.
It was found that the DWIA model gave an adequate description of these 
reactions when compared to data without including any damping mechanisms
for the delta resonance.
The model was also applied to eta photoproduction from
nuclei~\cite{Lee96}, 
and to kaon photoproduction from
nuclei~\cite{Lee97a} to investigate the possibility of using the
reaction to extract information about hyperon-nucleus interactions.

Section~\ref{dwia} outlines the key elements in the DWIA model.
Section~\ref{res} discusses some dynamical features of the 
basic production operator and reports our results on nuclei 
under quasifree kinematics.
Section~\ref{con} contains the concluding remarks.

\section{The DWIA Model}
\label{dwia}

Consider the process in the laboratory frame where the target is at rest.
The coordinate system is defined such that
the z-axis is along the photon direction ${\bf p_\gamma}$,
and the y-axis is along ${\bf p_\gamma} \times {\bf p_\pi}$ with 
the azimuthal angle of the pion chosen as $\phi_\pi = 0$.
The kinematics of the reaction are determined by
\begin{equation}
{\bf p_\gamma}={\bf p_\pi}+{\bf p_N}+{\bf p_m},
\label{mcon}
\end{equation}
\begin{equation}
E_\gamma +M_i = E_\pi +E_N +M_f + T_m.
\label{econ}
\end{equation}
Here ${\bf p_m}$ is the missing momentum in the reaction and 
$T_m={p_m^2 \over 2M_f}$ is the recoil kinetic energy.
The excitation energy of the residual nucleus is included in $M_f$.
The missing energy $E_m$ in the reaction is defined by 
$E_m=M_f-M_i+m_N=E_\gamma-E_\pi -E_N-T_m+m_N$ where 
$m_N$ is the mass of the nucleon.
For real photons, $|{\bf p_\gamma}|=E_\gamma$.
In the impulse approximation, 
the reaction is assumed to take place on a single bound nucleon whose 
momentum and energy are given by 
${\bf p}_i =-{\bf p_m}$ and $E_i=E_\pi +E_N -E_\gamma$.  
The struck nucleon is in general off its mass shell.  

The reaction is {\em quasifree}, meaning that the magnitude of 
${\bf p_m}$ has a wide range, including zero.
Since the reaction amplitude is proportional to the Fourier transform 
of the bound state single particle wavefunction, it falls off quickly 
as the momentum transfer increases. 
Thus we will restrict ourselves in the low $p_m$ region ($<$ 400 MeV/c)  
where the nuclear recoil effects ($T_m$) can be safely neglected 
for nuclei of $A > 6$.

The differential cross section cab be written as 
\begin{equation}
\frac{d^3 \sigma}{dE_\pi\,d\Omega_\pi\,d\Omega_N}=
{C \over 2(2J_i+1)}\;
\sum_{\alpha,\lambda,m_s}\frac{S_\alpha}{2(2j+1)}
|T(\alpha,\lambda,m_s)|^2.
\label{coin}
\end{equation}
The kinematical factor is given by
\begin{equation}
C=
{ M_f m_N\, |{\bf p_\pi}|\,|{\bf p_N}| \over 
4(2\pi)^5 
|E_N +M_f+T_m -E_N \,{\bf p_N}\cdot({\bf p_\gamma}-{\bf p_m})/p_N^2| }.
\end{equation}
The single particle matrix element is given by
\begin{equation}
T(\alpha,\lambda,m_s) = \int d^3 r\,
\Psi^{(+)}_{m_s}({\bf r},-{\bf p_N})\;
\phi^{(+)}_\pi({\bf r},-{\bf p_\pi})\;
t_{\gamma \pi}(\lambda, {\bf p_\gamma}, {\bf p_i} ,{\bf p_\pi}, {\bf p_N})\;
\Psi_{\alpha}({\bf r})\;
e^{i{\bf p_\gamma}\cdot{\bf r}}.  
\label{3d}
\end{equation}
In the above equations, $J_i$ is the target spin, 
$\alpha=\{nljm\}$ represents the single particle states,
$S_\alpha$ is called the spectroscopic factor,
$\lambda$ is the photon polarization, $m_s$ is
the spin projection of the outgoing nucleon, $\Psi^{(+)}_{m_s}$
and $\phi^{(+)}_\pi$ are the distorted wavefunctions with outgoing
boundary conditions, $\Psi_{\alpha}$ is the bound nucleon wavefunction,
and $t_{\gamma \pi}$ is the pion photoproduction operator.

In addition to cross sections, we also compute a polarization observable,
called photon asymmetry, defined by
\begin{equation}
\Sigma=
{ 
{d^3 \sigma}_{\perp}-{d^3 \sigma}_{\parallel}
\over
{d^3 \sigma}_{\perp}+{d^3 \sigma}_{\parallel}
}
\end{equation}
where $\perp$ and $\parallel$ denote the perpendicular and parallel
photon polarizations relative to the production plane (x-z plane).
We have used the short-hand notation
$d^3 \sigma \equiv d^3 \sigma/dE_\pi\,d\Omega_\pi d\Omega_N$ with 
appropriate sums over spin labels implied.
Note that the measurement of $\Sigma$ requires linearly 
polarized photon beams.

The dependence of the reaction on the nuclear structure is minimal.
It enters through the spectroscopic factor $S_\alpha$
and the single particle bound wavefunction.
The former is an overall normalization factor whose value can be 
taken from electron scattering. It cancels out in the photon asymmetry.
The latter can be sufficiently described by harmonic oscillator wavefunctions 
in the quasifree region we are interested in.

The final state interaction of the outgoing nucleon 
with the residual nucleus is described by optical potentials.
We use the global optical model by Cooper {\it et al}~\cite{Cooper93}
based on Dirac phenomenology that successfully describes the nucleon 
scattering data over a wide range of nuclei ($12\leq A \leq 208$) and 
energies (up to around $T_N=1000$ MeV).
To generate the distorted wavefunctions for our formalism, we 
numerically solved the Schr\"{o}dinger equation using the 
Schr\"{o}dinger equivalent potentials of the model.

For the outgoing pion distorted waves, we numerically 
solved the Klein-Gordon equation 
with the first-order optical potential by 
Gmitro {\it et al}~\cite{Gmitro87}. It is a microscopic global 
potential based on multiple scattering theory and supplemented by 
$\pi N$ phase shifts.
It gives an excellent description of 
the pion scattering data for nuclei $4\leq A \leq 40$
up to around $T_\pi=400$ MeV and valid up to $T_\pi=1000$ MeV.
We have checked that at low energies ($T_\pi \leq 220$ MeV), 
it is consistent with the phenomenological 
SMC pion optical potential~\cite{Stricker82} which was the traditional 
choice for low energy pions.

For the kinematics we will consider in the reaction, 
the kinetic energies of the pion and the nucleon vary in a wide range, 
up to about 900 MeV. Both are well covered by the optical potentials
considered. 
The integral in Eq.~(\ref{3d}) was evaluated numerically using 
techniques established in our earlier work~\cite{Lee93}.
To reach convergence, 10 to 15 partial waves were needed for the pion 
and around 20 partial waves for the nucleon.

All the dynamics of the photoproduction process is contained in the
elementary operator $t_{\gamma \pi}$ in Eq.~(\ref{3d}), 
which was usually given in terms of CGLN amplitudes.
To use it in a nuclear calculation,
we extrapolated the amplitudes off-shell
to account for the Fermi motion of the bound nucleon,
and performed necessary transformations into appropriate reference frames.
For our values of the momentum transfer,
the effects of the off-shell extrapolation were found to be quite small.

\section{Results}
\label{res}

\subsection{The elementary process}
\label{elem}

Recently the model of Ref.~\cite{Hanstein93}  was improved 
substantially by including more resonances and the new data 
from Mainz~\cite{Kamalov97}. 
The new model includes
the spin-1/2 resonances $P_{11}(1440)$ (the Roper resonance), 
$S_{11}(1535)$, $S_{11}(1650)$,
the spin-3/2 resonances $P_{33}(1232)$,
$D_{13}(1520)$, $D_{33}(1740)$,
and the spin-5/2 resonance $F_{15}(1680)$.
The resonances are described by Breit-Wigner parameterizations 
which have been unitarized by including energy-dependent phases that
have been adjusted to reproduce the appropriate multipole data.
The Born terms include the standard contributions from the nucleon, 
and the vector mesons $\rho$ and $\omega$.
This model is intended for photon energies up to about 1.1 GeV.
Among the considered resonances, the $D_{13}$ is the dominant one 
in the second resonance region (around $E_\gamma$=750 MeV), 
while the $F_{15}$ is the dominant one 
in the third resonance region (around $E_\gamma$=1000 MeV).

\begin{figure}
\centerline{\psfig{file=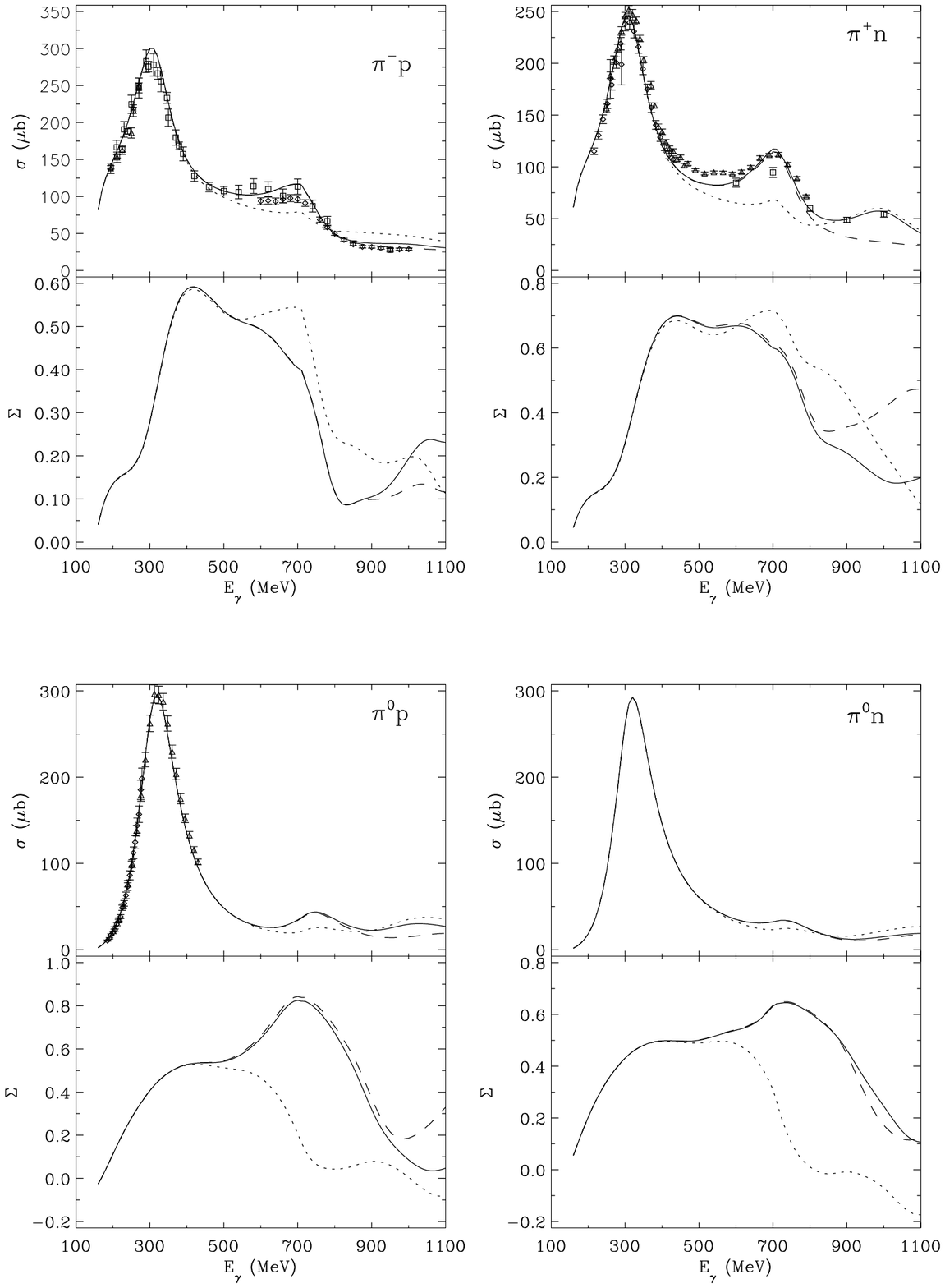,width=6in}}
\caption{Total cross section and photon asymmetry for the free process 
in the four channels.
The three curves correspond to results with the full operator (solid), 
with $D_{13}$ turned off (dotted), and with $F_{15}$ turned off (dashed). 
The data points are from~\protect\cite{SAID97} and~\protect\cite{Ukai85}.}
\label{tot3exp}
\end{figure}

To get an idea on the overall quality of the elementary operator, 
we first compare its predictions for the total cross section with 
available data in Fig.~\ref{tot3exp}.
Also shown are predictions for the photon asymmetry.
The three theoretical curves correspond to the full results (solid lines),
the results with the $D_{13}$ resonance turned off (dotted lines),
and with the $F_{15}$ resonance turned off (dashed lines).
Overall, the model gives a good description of the data.
The cross section results clearly show a second peak beyond the delta region,
more so in the charged pion channels than in the neutral pion channels,
which is dominated by the $D_{13}$ resonance.
A third peak is most apparent in the $\pi^+ n$ channel, 
due to the $F_{15}$ resonance.
It is interesting to observe that the photon asymmetry displays 
much larger sensitivities to the $D_{13}$ and $F_{15}$ resonances than 
the cross sections, especially to the $D_{13}$
resonance in the neutral pion channels.
Measurements of this polarization observable in the higher energy region 
will undoubtedly put more stringent tests on the basic ingredients 
in the elementary process. 

\subsection{Quasifree kinematics}
\label{res_qfree}

Now we turn to nuclear calculations for the reaction $A(\gamma,\pi\ N)A-1$.
As noted above, there is a great deal of kinematic flexibility in this
reaction.
Here we decide to present our calculations under quasifree kinematics.
It is achieved by solving Eq.~(\ref{mcon}) and Eq.~(\ref{econ})
at fixed $E_\gamma$ and $|{\bf p_m}|$ and $\theta_\pi$.
The quasifree kinematics closely resembles the two-body kinematics
in free space, except here it is on a bound nucleon with momentum $p_m$.
The energies are very close to those in free space 
(with small influences from nuclear binding and recoil).
The nucleon angle will be shifted from its free space value
by a certain amount depending on the value of $p_m$.
This kinematic arrangement has the feature that the 
variables vary in a wide range,
maximally exposing the underlying dynamics in the elementary operator,
while being minimally sensitive to the details of the nuclear wavefunction.  
It benefits studies of the final state interactions as well as the 
elementary process.

We will limit ourselves to coplanar set-ups with the nucleon on the 
opposite side of the pion ($\phi_N=180^0$). Such set-ups generally result in 
larger cross sections than out-of-plane set-ups.
We will use $^{12}C$ as an example. Other nuclei could be studied 
straightforwardly.
Specifically, we will present results for the reaction 
$^{12}C(\gamma,\pi N)X_{g.s.}$ 
(the final nucleus $X$ is either $^{11}C$ or $^{11}B$ left in its ground state) with $p_m$=100 MeV/c. 
This value of $p_m$ yields large counting rates for p-shell nuclei 
since a 1p-shell nucleon wavefunction has its maximum value at roughly 
that value.
As an example, we give in Table~\ref{kin} the numerical solutions for the 
kinematics at $E_\gamma=750$ MeV
for this reaction in the $\pi^- p$ channel.
Small differences occur in the other three channels due to small 
mass differences of the particles.
Both the pion and the nucleon are more energetic in the forward directions.
The nucleon tends to stay in the forward direction due to its heavier mass.
These numbers also demonstrate the influence of the nuclear binding 
effects as discussed above.

\begin{table}[tb]
\caption{Quasifree kinematics for the reaction 
$^{12}C(\gamma,\pi^- p)^{11}C_{g.s.}$ at $E_\gamma=750$ MeV 
in the laboratory frame.
The numbers in parentheses are the corresponding solutions 
on the free nucleon.}
\label{kin}
\begin{tabular}{cccc}
$\theta_\pi$ (deg) & $T_\pi$ (MeV)  & $T_p$ (MeV) & $\theta_p$  (deg) 
\\ \hline
30  &  520.8 (538.2)  &  72.2 (72.3)    &  74.7 (72.3) \\
60  &  382.9 (398.4)  &  210.1 (212.0)  &  49.9 (42.4) \\
90  &  269.9 (283.0)  &  323.1 (327.4)  &  34.0 (27.9) \\
120 &  200.3 (211.5)  &  392.7 (399.0)  &  22.6 (16.9) \\
150 &  164.8 (174.7)  &  428.2 (435.7)  &  13.6 (8.0 ) \\
\end{tabular}
\end{table}

\begin{figure}
\centerline{\psfig{file=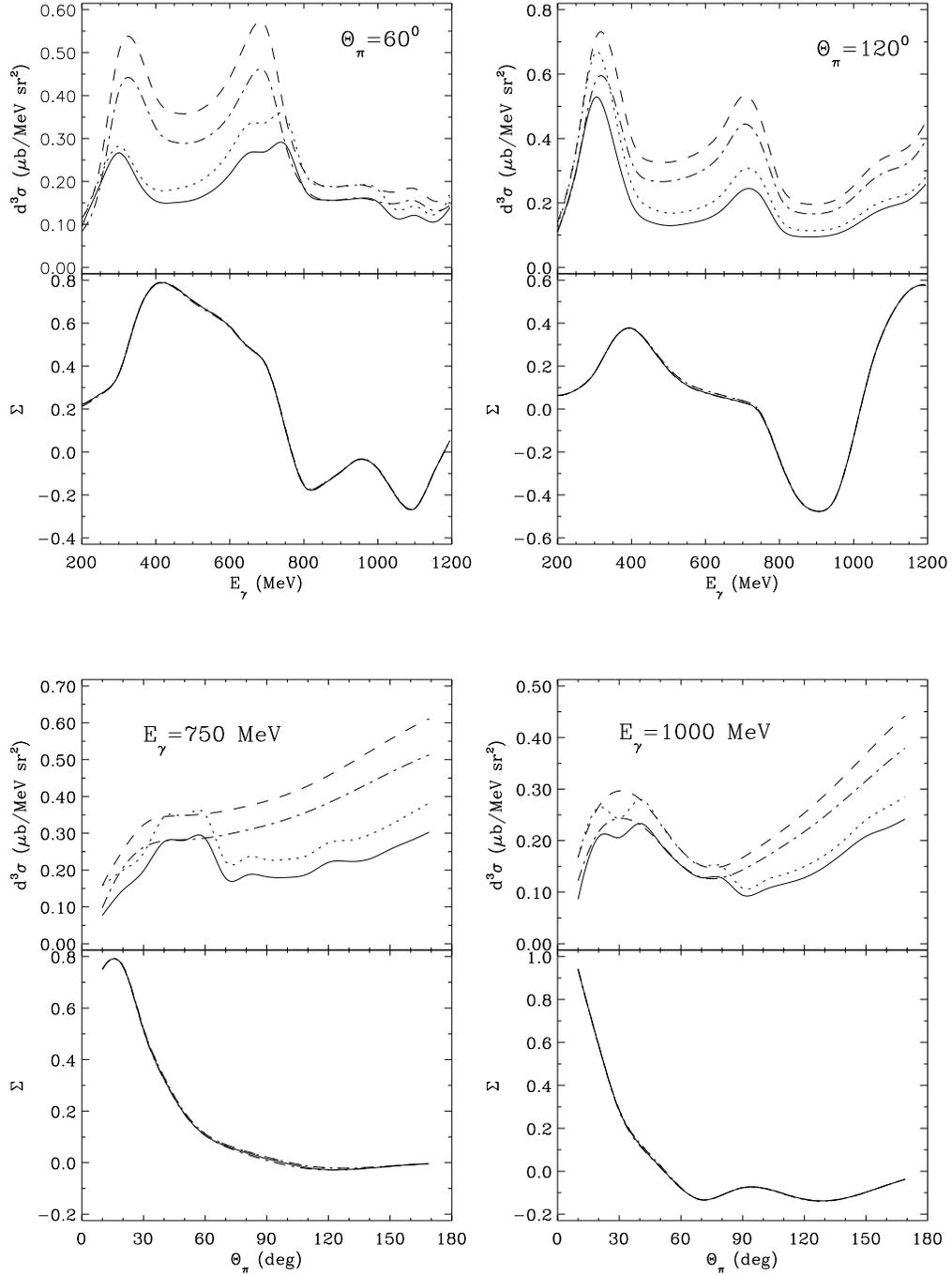,width=5.9in}}
\caption{Effects of final state interactions in the reaction
$^{12}C(\gamma,\pi^- p)^{11}C_{g.s.}$.
The upper two panels are photon energy distributions 
at fixed pion laboratory angles,
while the lower two are pion angular distributions at fixed photon energies.
The four curves correspond to calculations in PWIA (dashed), 
in DWIA with pion distortion only (dotted), 
with nucleon distortion only (dash-dotted), and with both distortions (solid).}
\label{dist1}
\end{figure}

An important aspect of the reaction is the final state 
interaction (FSI) of the 
exiting pion and nucleon with the residual nucleus. They are 
accounted for by optical models taken from elastic scattering. 
Here the reaction also serves as a testing ground of these models 
in a wide energy range. 
Fig.~\ref{dist1} shows the effects of final state interactions in the
$\pi^- p$ channel.
To ensure wide kinematical coverage,
the total and individual FSI effects on the coincidence cross
section ($d^3\sigma$) and the photon asymmetry ($\Sigma$) 
are shown in two distributions: as a function of 
$E_\gamma$ at two fixed values of $\theta_\pi$, and as a function of
$\theta_\pi$ at two fixed values of $E_\gamma$.

First we examine the cross sections.
The effects of the nucleon FSI alone are relatively small,
causing almost an overall reduction of less than 20\%.
The effects of the pion FSI alone are more significant.
In the energy distributions at $\theta_\pi=60^0$, 
they are large (about a factor of 2) below $E_\gamma$=800 MeV,
and become quite small after that. This behavior is consistent 
with the fact from elastic scattering that 
pion distortion decreases with increasing energy.
The $T_\pi$ in this case grows from about 40 MeV to 400 MeV as 
$E_\gamma$ increases from 300 MeV to 1100 MeV.
At $\theta_\pi=120^0$, $T_\pi$ changes from about 80 MeV to 690 MeV.
In the angular distributions, pion distortion is more significant 
at backward angles (up to a factor of 2 reduction) 
than at forward angles (almost no reduction), and 
results in a non-smooth angular distribution.
We have checked that the results are convergent by varying the number of 
partial waves and integration points.
To further check this behavior, we repeated the same calculation 
by setting the elementary operator to unity and found that the 
resulting distribution was rather smooth.
This suggests that the non-smooth behavior was due to the interference 
between the pion FSI and the elementary process. 
As a result, the total distortion largely follows the shapes of the 
pion distortion with additional reduction due to the nucleon distortion.

As for the photon asymmetry, 
the inclusion of FSI results in  almost no changes in this observable in 
all instances.  This situation is similar to our previous findings 
in quasifree pion and eta photoproduction~\cite{Lee93,Lee96,Lee97}.
Furthermore, the photon asymmetry is not sensitive to the nuclear 
structure input as alluded to earlier.
We have examined FSI effects in the other three channels, and found 
similar effects which will not be shown here.

The insensitivity of the photon asymmetry to FSI and to nuclear structure
opens the way to use this observable to study 
possible modifications of the basic production process in the nuclear medium.
Such modifications may manifest themselves through variations in the 
parameters of the resonance such as the mass, the width and the coupling 
constant (strength).
It is expected that its mass would be decreased from interactions 
via a nuclear potential.
Additional channels might open in the medium, 
causing an increase in its decay width.
The coupling at the interaction vertices might also be modified.
To perform a phenomenological sensitivity study of the 
$D_{13}$ and $F_{15}$ resonances,
we apply a 3\% decrease in their masses and a 60\% broadening of their
widths in the elementary operator.  
These values are in line with those suggested by an 
cooperative damping model of the resonances~\cite{Hirata97}.
As for the strength, we use the values suggested by a quark 
exchange model~\cite{Giannini94}: a reduction of the $D_{13}$ strength
by 12\% and the $F_{15}$ strength by 22\%.
We do not claim any validity of these models, but are mostly interested
in exploring sensitivity of the observables to possible changes in the
resonance parameters.
Rigorous studies of such effects clearly require a more dynamical treatment,
as in the delta-hole model.

\begin{figure}
\centerline{\psfig{file=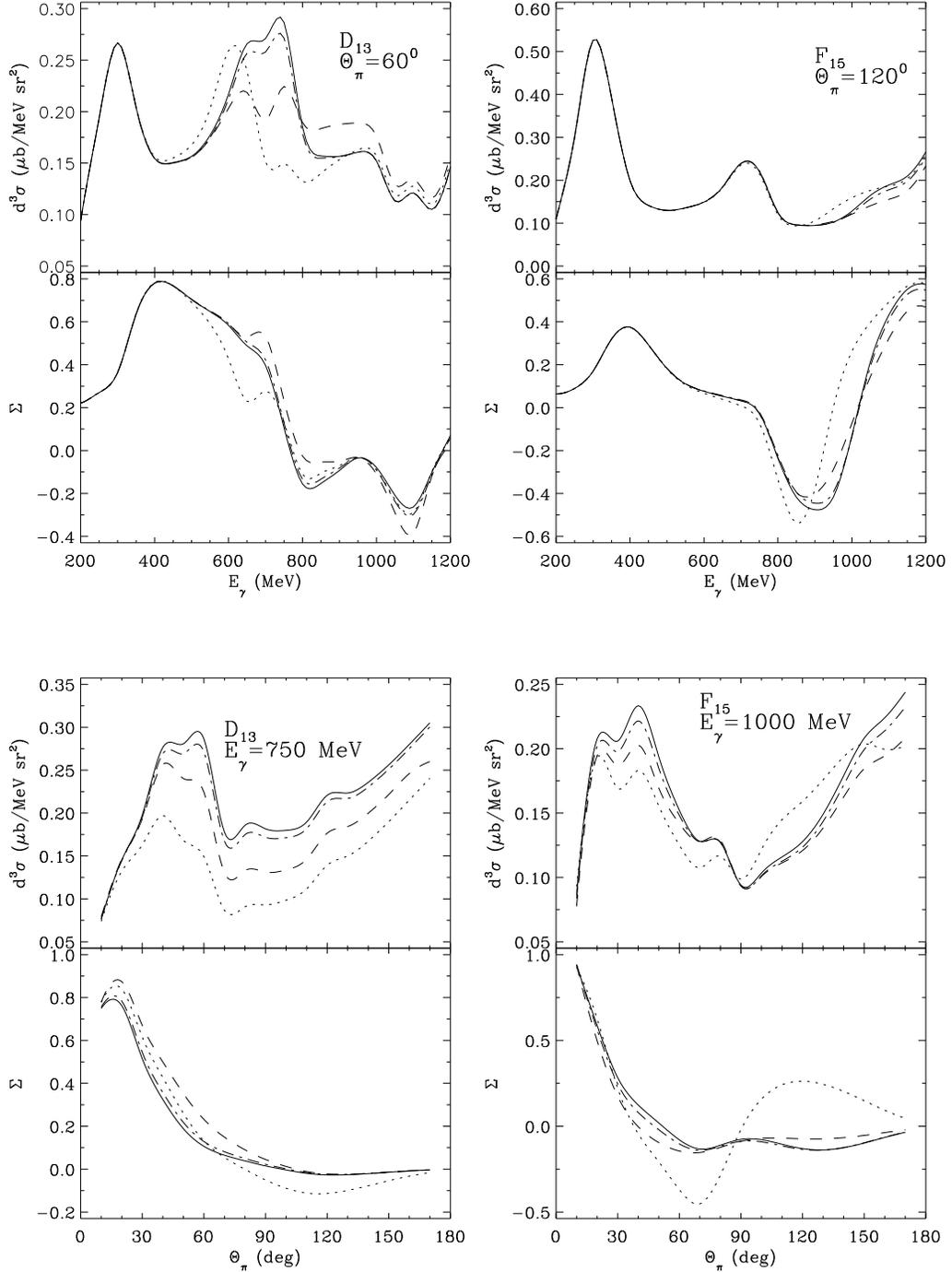,width=6in}}
\caption{Effects of possible medium modification of the $D_{13}(1520)$ 
and $F_{15}(1680)$ resonances in the reaction 
$^{12}C(\gamma,\pi^- p)^{11}C_{g.s.}$.
The four curves correspond to the full results (solid), 
with the mass decreased by 3\% (dotted),
with the width increased by 60\% (dashed),
and with the strength decreased by 12\% for $D_{13}$ and 22\% for $F_{15}$
(dot-dashed).
The calculations were done in DWIA.}
\label{med1-dw}
\end{figure}

\begin{figure}
\centerline{\psfig{file=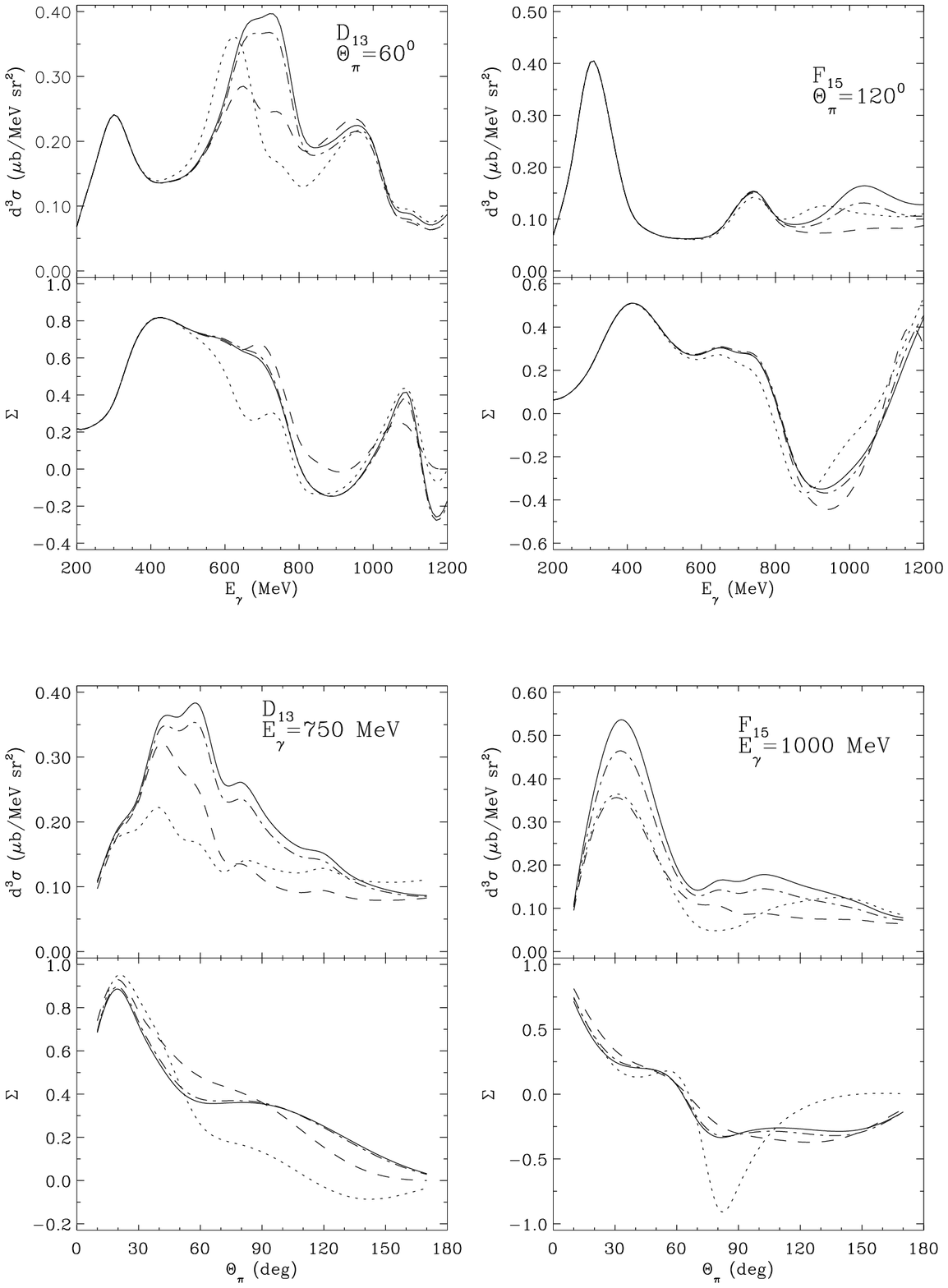,width=6in}}
\caption{Similar to Fig.~\protect\ref{med1-dw}, 
but for the $\pi^+ n$ channel.}
\label{med2-dw}
\end{figure}

\begin{figure}
\centerline{\psfig{file=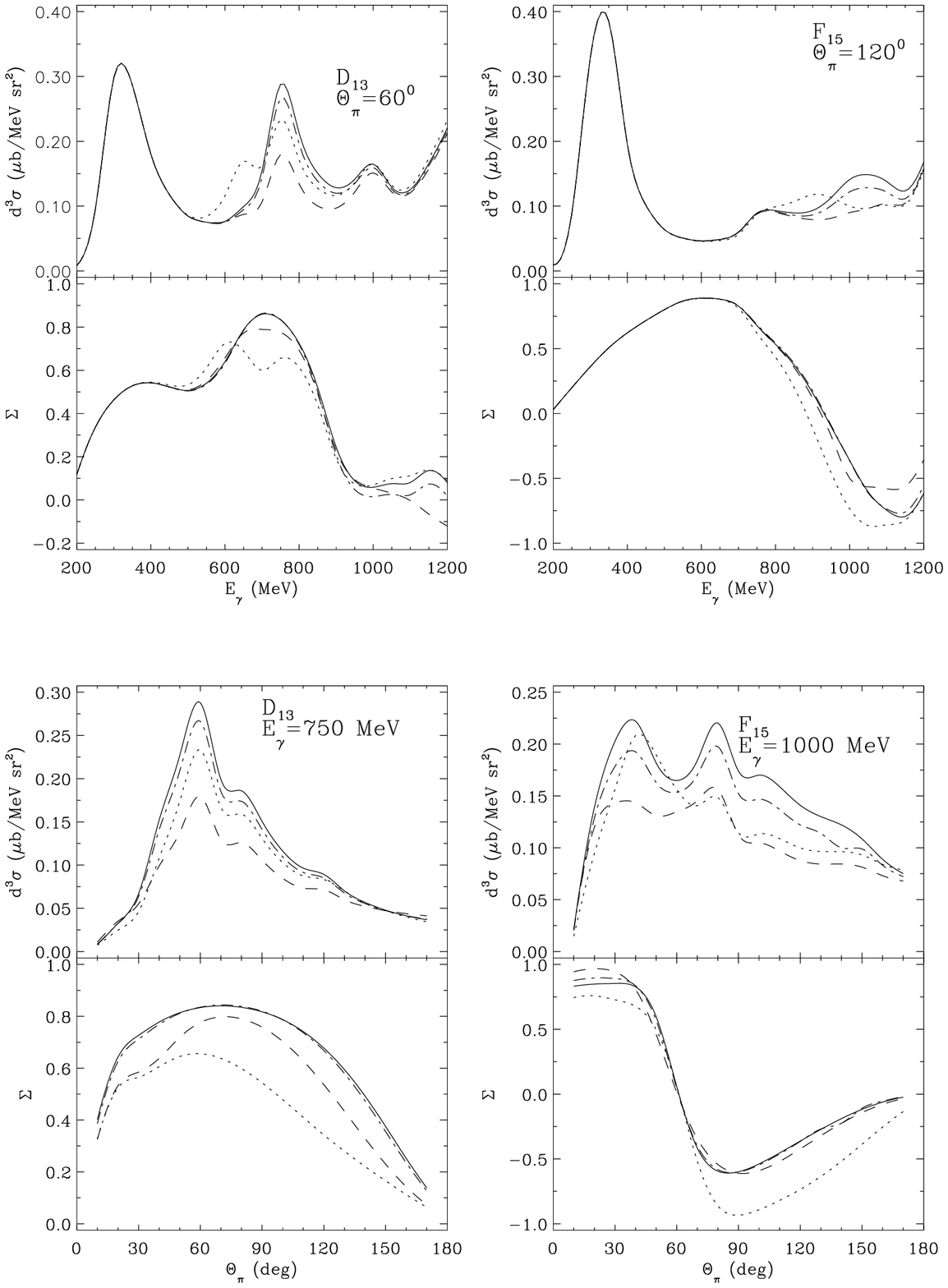,width=6in}}
\caption{Similar to Fig.~\protect\ref{med1-dw}, 
but for the $\pi^0 p$ channel.}
\label{med3-dw}
\end{figure}

\begin{figure}
\centerline{\psfig{file=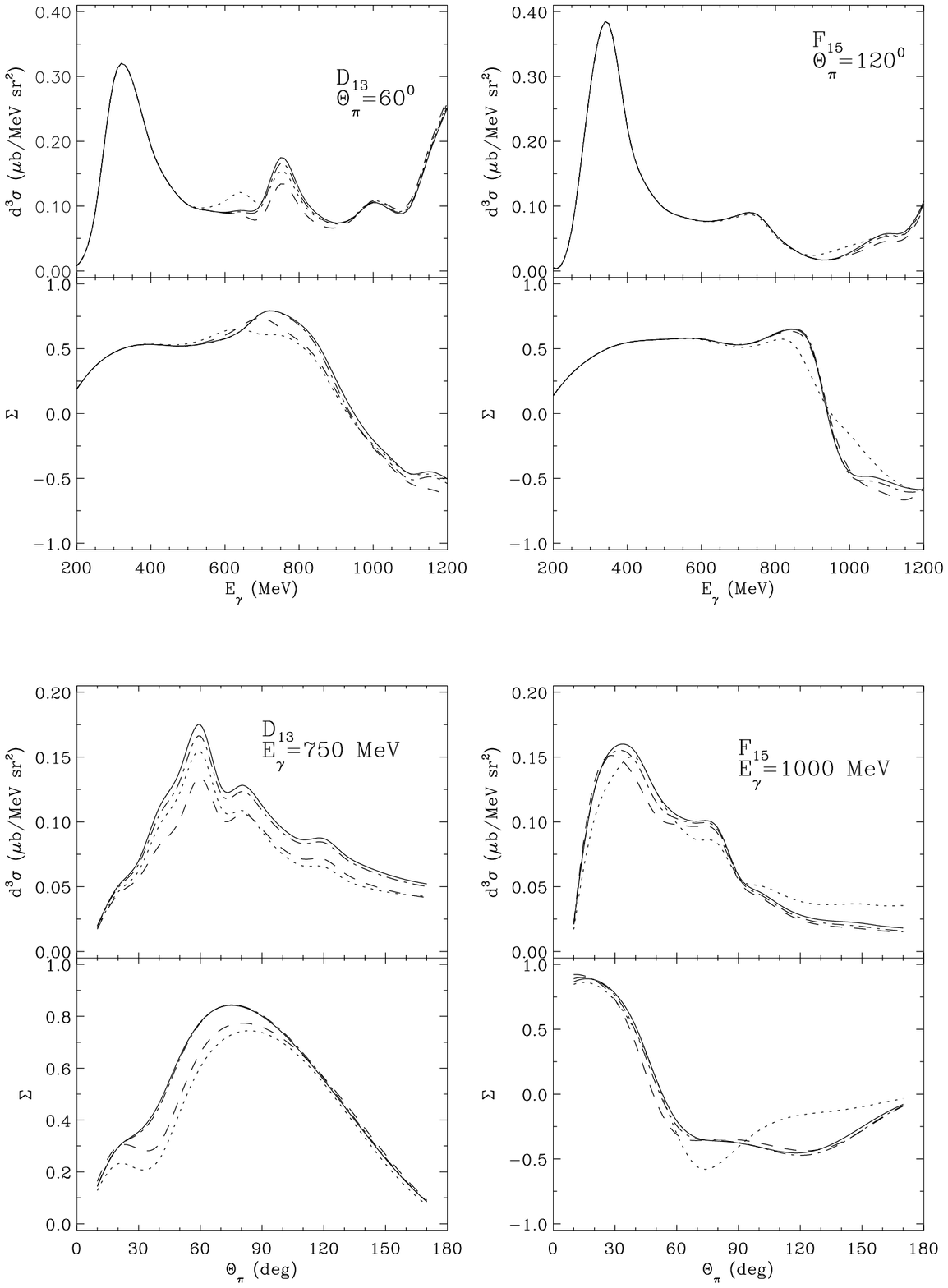,width=6in}}
\caption{Similar to Fig.~\protect\ref{med1-dw}, 
but for the $\pi^0 n$ channel.}
\label{med4-dw}
\end{figure}

In Figs.~\ref{med1-dw} to~\ref{med4-dw}, 
we present the possible medium effects 
at kinematics similar to Fig.~\ref{dist1}.
The energy distribution for the $D_{13}$ is shown at a forward angle 
of $\theta_\pi=60^0$, while for the $F_{15}$ it is shown at
$\theta_\pi=120^0$.
The angular distributions are shown at $E_\gamma$ sitting 
on top of the resonances, at 750 MeV and 1000 MeV.
Fig.~\ref{med1-dw} shows the effects of medium modifications 
for the process $^{12}C(\gamma,\pi^- p)^{11}C_{g.s.}$
The four curves correspond to the full results (solid),
the results with the mass decreased by 3\% (dotted),
with the width increased by 60\% (dashed),
and with the strength decreased by 12\% for $D_{13}$ and 22\% for
$F_{15}$ (dot-dashed).  
The results are most sensitive to changes in the resonance mass,
followed by the resonance width.
The sensitivity to the resonance strength is surprisingly small.
In the cross sections, the small reduction of the $D_{13}$ mass results 
in a large shift of the second peak as well some reduction 
in the energy distributions, and a large reduction in the angular
distributions.  The effects of the change in width are also quite large.
As for the `cleaner' photon asymmetry, there is a noticeable effect due to the 
mass change for the $D_{13}$ in the energy distribution, 
and a very large sensitivity to the $F_{15}$ in the angular distribution. 
These effects are large enough to be detected by experiments.

Fig.~\ref{med2-dw} shows the same effects in the $\pi^+ n$ channel.
Here all changes in the $F_{15}$ act to damp out the third peak in the 
energy distribution.  Similar sensitivities to the $D_{13}$ are observed. 
For completeness, the same distributions are shown in 
Fig.~\ref{med3-dw} for the $\pi^0 p$ channel,
and in Fig.~\ref{med4-dw} for the $\pi^0 n$ channel.
It should be pointed out that these calculations were done in DWIA, 
so the effects in the cross section 
are {\it relative} effects on top of the FSI effects.
The effects in the photon asymmetry are almost purely signatures of 
medium modifications, as discussed earlier.

\section{Conclusion} 
\label{con}

	Using a tree-level model for single pion photoproduction 
on the nucleon at higher energies, 
we have performed calculations for the exclusive photonuclear reaction, 
$A(\gamma,\pi N)A-1$, 
in the distorted wave impulse approximation,
for photon energies up to about 1.1 GeV.
The reaction has the kinematical feature that it 
allows small momentum transfers.
In the quasifree region ($<$ 400 MeV/c), the reaction is not 
sensitive to the nuclear structure input, and provides one of the 
cleanest ways to study nucleon resonances in the nuclear medium. 

Final state interactions are incorporated via optical potentials and
are found to substantially reduce the cross sections, especially at
backward pion angles, but not affect the photon asymmetry.

To shed light on the `damped resonances' in the second and third 
resonance regions as seen in total nuclear photoabsorption data,
we explored the effects of possible medium modifications of the 
$D_{13}$ and $F_{15}$ resonances by varying their parameters 
using model guidance. 
It is found that this reaction is very sensitive to such changes in 
these resonances. 
Especially if the photon asymmetry is measured, the effects can 
easily be detected by experiments.
We therefore believe that measurements of exclusive quasifree pion 
photoproduction with polarized photons are an important future step 
that should help untangle the riddle of the damped resonances.

\acknowledgements
This work was supported in part by U.S. DOE under Grants 
DE-FG03-93DR-40774 (F.L.),
DE-FG02-95ER-40907 (C.B.),
 and DE-FG02-87ER-40370 (L.W.).

\end{document}